# Effect of COVID-19 Pandemic on Oceans

Maryam R. Al Shehhi*

*Civil, Infrastructural and Environmental Engineering Department, Khalifa University, Abu Dhabi, UAE*

Maryam R. Al Shehhi (maryamr.alshehhi@ku.ac.ae) the *corresponding author

# Effect of COVID-19 Pandemic on Oceans

The novel corona virus (COVID-19) has slowed down a lot of human activities in the world. A lockdown resulting in the closure of largest industries in the world for a period of 2 months, due to the pandemic, was enough to cause a drop of 7% of the anthropogonic $CO_2$ in the atmosphere. In addition to the world in general, the excess of the anthropogonic $CO_2$ emission in the atmosphere has always been a threat to the oceans as well. Oceans play a key role to buffer the greenhouse effect, but in the process, it becomes warmer, more acidic, and less oxygenated. While there have already been investigations done on the effect of pandemic on atmosphere, the question what happens to oceans during the pandemic remains unanswered. The aim of this paper is to study the effect of the pandemic and the resultant reduction in $CO_2$ emissions on the productivity of the global oceans. Often Chlorophyll-a (Chl-*a*), Particulate organic and inorganic carbon (PIC: POC), and sea surface temperature (SST), are used to indicate the productivity of oceans. Herein, satellite-derived estimates of the aforementioned parameters are used. Based on these estimates, a drop in Chl-*a* (≥0.5 mg.m$^{-3}$) is observed off Alaska, North Europe, South China and Southeast USA during the pandemic. $CO_2$ reduction of 123 metric ton of $CO_2$ (MtCO$_2$) during the pandemic in China might have caused reduction in mean Chl-*a* by around 5% (2.5 to 1.6 mg.m$^{-3}$). Reduction of Chl-*a* during the pandemic is mostly associated with the reduction of PIC:POC. The pandemic demonstrates noticeable effect on Chl-*a* and/or SST. A cooling response of 0.5 $^o$C in mean SST is observed over most of the coastal areas, especially off Alaska, north Indian ocean, and eastern Pacific. The decrease in the $CO_2$ emissions in India by 30% during the pandemic translates into a drop of mean SST in the north Indian ocean by 5% (from 29.95 to 28.46 $^o$C). All these suggest that maintaining global activities as sustainable as the pandemic period, can help to recover the oceans.

## 1. Introduction

Since the pandemic has erupted there have been a lot of discussions about the effect of pandemic on the environment, pollution, economy, employment and what not. But, what about the oceans?

Oceans are the largest source of food, with more than 40% of the world's population relying on the oceans as their primary source (UN SDG 2016). Oceans are also important in regulating the global climate in terms of balancing the temperature of earth (Morel and Antoine 1994), simulating the rainfall (Goddard and Graham 1999) and providing 50-80% of the oxygen in earth (NOAA 2020). Therefore, any change on the surface of the ocean has a direct effect on the life on the earth.

As the pandemic has shown a pronounced effect on the environment in general (Zambrano-Monserrate, Ruano, and Sanchez-Alcalde 2020), it is imperative to study its effect on oceans. The characteristic of the oceans, which directly influence the life on the earth is their productivity. Productivity represents the health of the marine ecosystem and the carbon cycle (Behrenfeld et al. 2005). Productivity is commonly estimated as the plant biomass in the ocean and Chlorophyll-*a* (Chl-*a*) is one of the key metric indicators for it (Morin, Lamoureux, and Busnarda 1999). Ocean productivity is associated with the sea temperature (Behrenfeld et al. 2006; Gerecht et al. 2014) and the carbon cycle. The carbon cycle is affected by two main carbon processes including the production of particulate organic carbon (POC) during biomass photosynthesis and the formation of carbonate shells (PIC) during biomineralization (Gehlen et al. 2006). The latter is an important element of global ocean carbon, which consists of calcium carbonate shells. These shells partially dissolve in the oceans and what remain deposit on the seafloor of oceans (Mitchell et al. 2017).

The COVID-19 pandemic has slowed down a lot of human stressors in the world (Le Quéré et al. 2020). The pandemic erupted in November 2019. However, it has just been over 2 months for the closure and lockdowns of big industries such as the automobile factories in China (e.g. General Motors, Honda Motor, Nissan Motor), Europe and Americas (e.g. General Motors, Ford Motor)(UN ILO 2020b), textile and clothing factories of the major brands, including Adidas, Gap, H&M and Inditex (UN ILO 2020c), and the maritime fishing and shipping industry where many fishing vessels are unable to leave port and the demand for many seafood products is substantially reduced (UN ILO 2020a). All these human stressors have been seriously affecting the ocean in the past decades until the pandemic has occurred. This effect includes the high sea surface temperature (SST) (Knutson et al. 2010; Nurhati, Cobb, and Di Lorenzo 2011; Yeh et al. 2009), ocean acidification (Boyd 2011; Ekstrom et al. 2015; Wittmann and Pörtner 2013) and increasing ultraviolet radiation (Williamson et al. 2014). With continued business-as-usual emissions, SST is projected to increase 0.035 °C per year and warm an additional 2.8 °C by 2100 (Bruno et al. 2018). There is no answer to the question whether or not the pandemic has helped in reducing the cumulative human impact on the oceans through the reduction of 7% in the global $CO_2$ emissions (Le Quéré et al. 2020).

This study is dedicated to answer this question and to study the pandemic's effect on other parts of our world. The analysis done herein is based on the ocean's data retrieved from the satellite images during the pandemic and pre-pandemic periods. The changes of the Chl-*a*, SST, PIC:POC are addressed during these periods in global and regional perspectives.

**2. Results**

In this analysis, the daily level-3 ocean data retrieved from the Moderate Resolution Imaging Spectroradiometer (MODIS) are used. The data were obtained with a resolution of 9-km for the period between 6$^{th}$ of April and 15$^{th}$ of Jun (2019 and 2020). This period has been selected to study the changes of the surface properties of the oceans during the pandemic of 2020 compared to those of the 2019 to provide a quantitative measure of relative change compared to pre-COVID conditions.

To particularly study the changes of the global ocean productivity during the pandemic, this work is focusing on analyzing the changes of chlorophyll-a (Chl-*a*: $mg.m^{-3}$) and sea surface temperature (SST: $^oC$) retrieved from MODIS (Polovina, Howell, and Abecassis 2008; Saba et al. 2010; Roxy et al. 2016; Goes et al. 2000). Whereas, changes of the ratio of MODIS PIC ($mg.m^{-3}$) to POC ($mg.m^{-3}$) are used to study the changes in the carbon fluxes (Balch et al. 2007; Lipsen et al. 2007; Findlay, Calosi, and Crawfurd 2011). The difference of ten-day average of MODIS Chl-*a*, SST, PIC:POC maps have been calculated to analyze the global changes of ocean productivity and carbon fluxes for May and Jun during the pandemic and pre- pandemic periods. In addition, daily MODIS Chl-*a*, SST, PIC:POC are used to analyze the time series of these parameters during the period between 6$^{th}$ of April and 15$^{th}$ of Jun in both the pre-pandemic period (2019) and the pandemic period (2020). These time series data are obtained for 11 regions: *Alaska*, *Northeast United States (USA)*, *Southeast USA*, *Pacific Ocean*, *Southeast America*, *China & South Korea*, *Middle East*, *North Europe*, *Northwest Africa*, *Southwest Africa* and *Southeast Australia*. These regions are selected because of the presence of either high industrial activities (UNIDO 2019) or large population (WHO 2016). The pandemic effect, if any, is expected to be obvious in these intense industrial and/or highly populated areas. Therefore, to understand the pandemic effect on the productivity of oceans analysis has been performed in global and regional perspectives.

*Global Productivity Changes*

The global Chl-*a* concentrations of 2020 are subtracted from those of the 2019 as shown in Figure 1a. Based on this difference in Chl-*a* (2020-2019), there is mostly a drop of Chl-*a* in the northeast USA in 2020 compared to 2019 during May and Jun. The drop of Chl-*a* has been also observed in the north Europe in 2020 compared to 2019, especially during May. There is also a drop of more than 2 $mg.m^{-3}$ in southeast America, along the Argentine basin which always experience high Chl-

*5*

*a* concentrations due to the high surface eddy kinetic energy that brings up the nutrients in the mixed layer (Richardson, Weatherly, and Gardner 1993). The drop of Chl-*a* can be also observed in the south coasts of China. This drop of Chl-*a* could be very much caused by the reduction of the $CO_2$ emissions during the pandemic period (Denman 2008). Because phytoplankton biomass takes up the atmospheric CO2 through the photosynthetic primary production for their growth. Therefore, as seen here, the reported $CO_2$ reduction of 123 metric ton of $CO_2$ ($MtCO_2$) during the pandemic (Le Quéré et al. 2020) in China has caused reduction in the mean Chl-*a* by around 5% (from 2.5 to 1.6 mg.m$^{-3}$).

The reduction in atmospheric $CO_2$ emissions during the lockdowns have also affected the carbon cycle in the ocean based on the changes observed in the ratio of PIC:POC as evident in Figure 1b. The ratio of PIC to POC (PIC:POC) is well representing the carbon cycle and the pumping of carbon in to the deep oceans. Because the $CO_2$ consumed during photosynthetic primary production is converted to the organic molecules with some species also forming $CaCO_3$ called PIC. All these particulate ecosystem carbon, living and dead, is grouped together as POC plus PIC (Denman 2008). Therefore a decrease in PIC:POC indicates a reduction in $CO_2$ uptake. The changes of PIC:POC is observed based on the subtraction of PIC:POC values of 2020 from those of the 2019 during both May and Jun (Figure 1b). As shown in these maps, generally, there is a reduction of PIC:POC off the global coasts. For example, there is a reduction of around 0.5 (0.6-0.14) in PIC:POC off Europe during Jun. This means that the reported reduction of atmospheric $CO_2$ emissions by 24% (Le Quéré et al. 2020) during the pandemic in Europe can cause a direct decrease of PIC:POC by 75%.

The reduction in $CO_2$ emission caused by the pandemic has also caused a cooling response of 0.5 ºC in SST over most of the coastal areas, especially off Alaska, north Indian ocean, and eastern

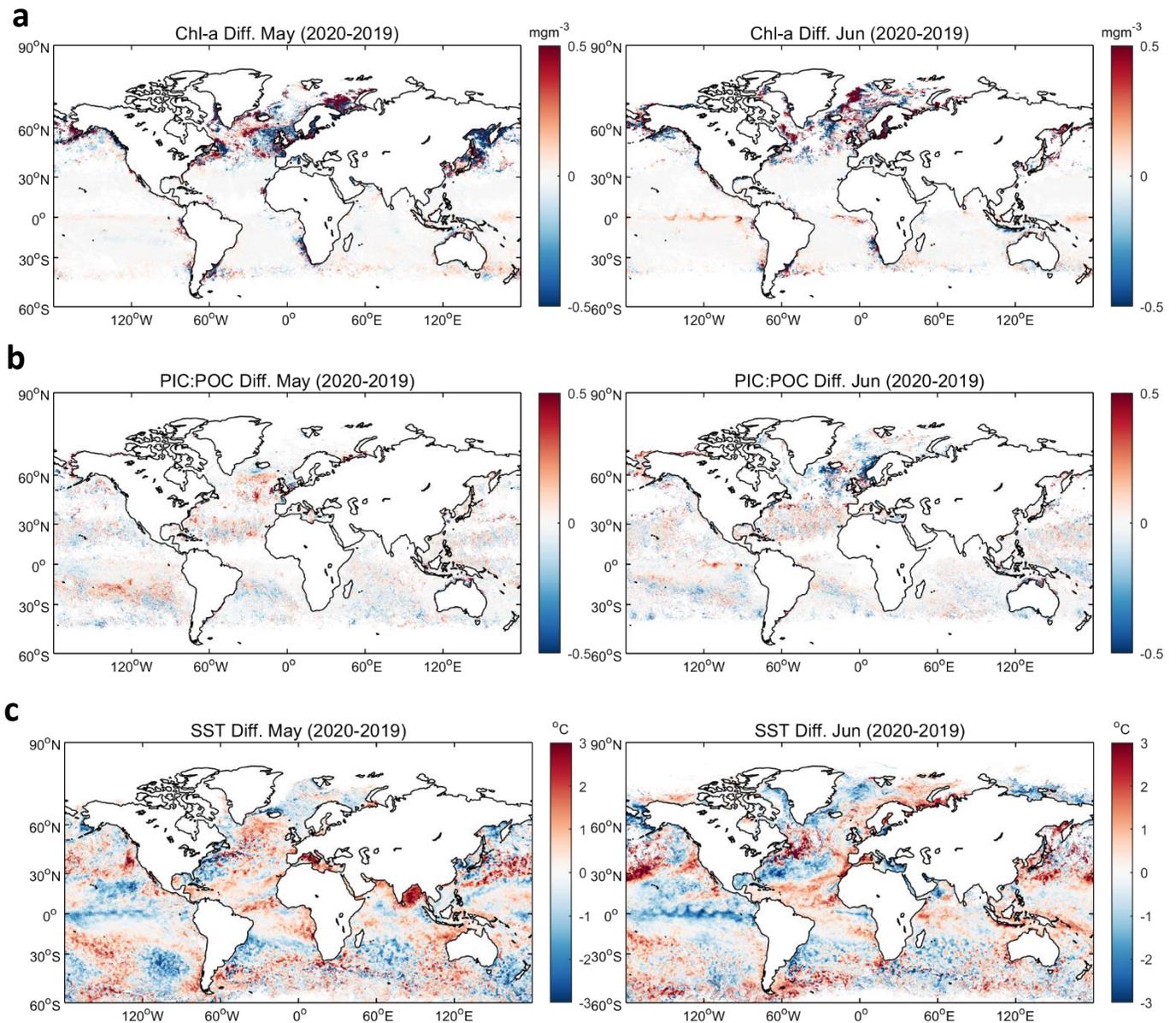

Figure 1. Maps of 10-days average maps for **a** Chl-*a* (mg.m$^{-3}$) **b** PIC:POC & **c** SST (°C) difference between the pandemic period including Jun and May and for those of the pre-pandemic period (2019). The blue colour represents the negative change that refers to reduction in the corresponding parameter in each map. The red colour represents the positive change that refers to reduction in the corresponding parameter in each map.

Pacific as shown in the figure of SST difference between 2020 and 2019, Figure 1c. These regions have been reported to have high warming trend previously caused by climate change (Reid et al. 2010). This can explain their fast response to the reduction in $CO_2$ emissions. The decrease in the $CO_2$ emissions in India by 30% during the pandemic might have caused a drop of SST in the north

Indian ocean by 5 % (29.95-28.46 °C) during the month of Jun of 2020. By comparing the SST and Chl-*a* maps, a reduction trend in SST and an increase of Chl-*a* is observed. This is because the reduction in SST can improve the uptake of the atmospheric $CO_2$ by the ocean and can enhance the photosynthesis process. Therefore, the reduction in $CO_2$ emissions doesn't have a direct effect on Chl-*a* and SST, rather it is related with both of them.

*Regional Productivity Changes*

Based on the aforementioned analysis of the pandemic effect on the global oceans, 11 regions have been identified to be more critical than others as per the effect of pandemic. Therefore, the daily time series of Chl-*a*, PIC:POC, and SST have been obtained for these 11 regions starting from 6$^{th}$ of April until 15$^{th}$ of Jun for 2019 and 2020.

Daily time series data suggest that there is a general decrease of Chl-*a* during the pandemic in most of the regions especially in China & South Korea, southeast America, Pacific Ocean, Southwest Africa, Alaska, and Middle East compared to 2019 as shown in Figures 2a-2f. For example, the mean Chl-*a* concentrations along Southwest Africa have dropped during the pandemic period of Jun from 2.5 mg.m$^{-3}$ to 2 mg.m$^{-3}$ compared to Jun 2019 (Figure 2d). Likewise, there is a prominent drop of Chl-*a* from 3 mg.m$^{-3}$ to 2.5 mg.m$^{-3}$ during the pandemic period off Alaska region as shown in Figure 2e. A reduction of 9% (from 2.2 to 1.07 mg.mg$^{-3}$) in Chl-*a* is observed in the Middle East (Arabian Gulf) as evident in Figure 2f. In addition, in the Southeast America, for example, on 28$^{th}$ of May, Chl-*a* decreased by 3 mg.m$^{-3}$ in 2020 compared to 2019 (Figure 2b). The drop of Chl-*a* has reached up to 50% during the pandemic in Alaska, Southeast America, Pacific Ocean, and Southwest Africa. In some of the regions, the reduction in $CO_2$ emissions have reduced the degree of Chl-*a* fluctuations in 2020 compared to 2019. For example, in Southwest America, North Europe, and Pacific Ocean the maximum peaks have dropped during the pandemic by 28%, 40%

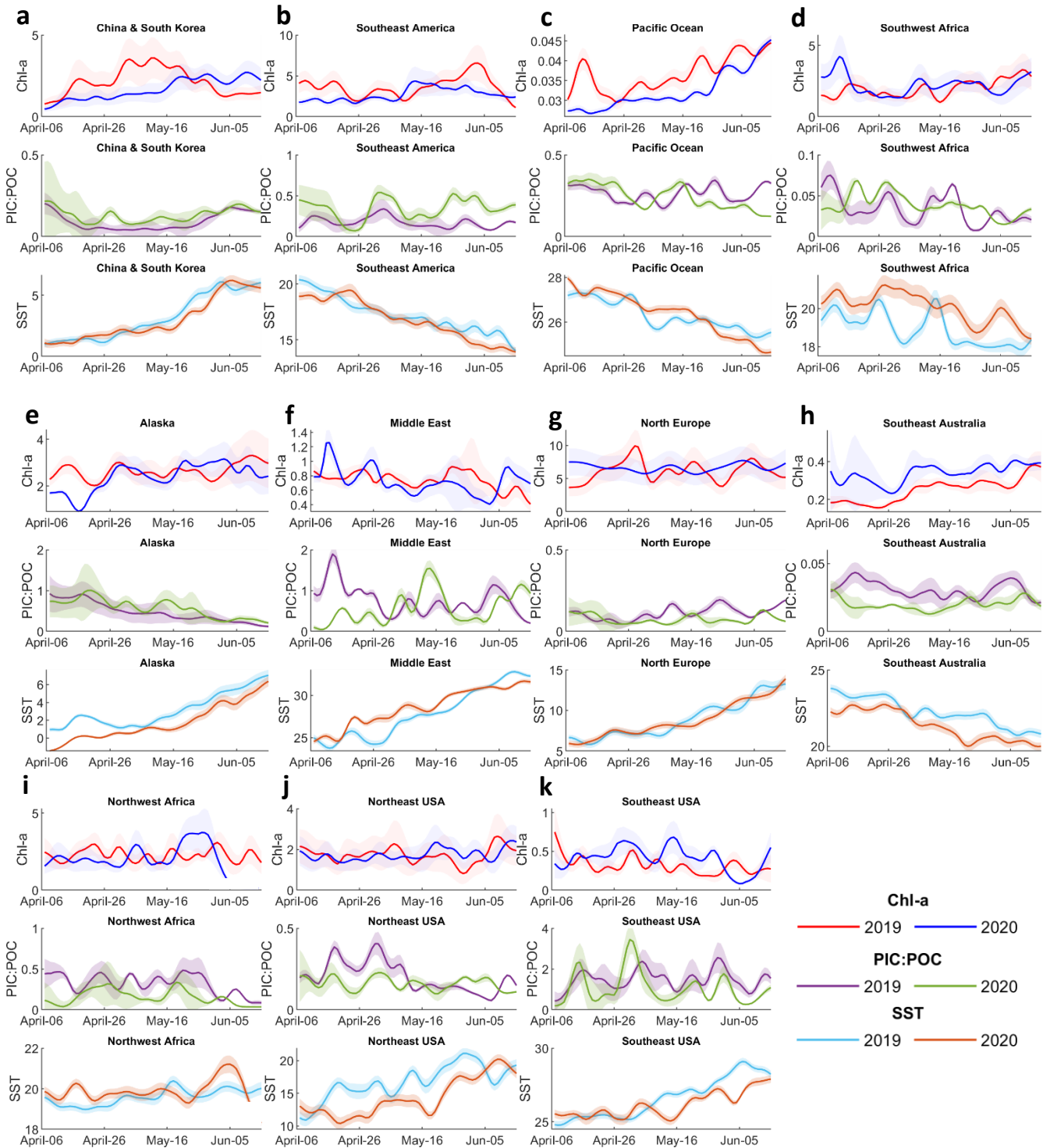

Figure 2. Daily time series data of Chl-*a* (mg.m$^{-3}$), PIC:POC and SST (°C) during the pandemic period (April -Jun) 2020 and pre-pandemic period (April-Jun) 2019 for 11 regions **a** China & South Korea **b** Southeast America **c** Pacific Ocean **d** Southwest Africa **e** Alaska **f** Middle East **g** North Europe **h** Southeast Australia **i** Northwest Africa **j** Northeast USA **k** Southeast USA. In the Chl-*a* time series, the blue line represents 2020 and the red line represents 2019. In the PIC:POC time series, the green line represents 2020 and the purple line represents 2019. In the SST time series, the orange line represents 2020 and the light blue line represents 2019. All of these bold time series lines represent the mean values with 95% confidence interval.

*9*

and 12%, respectively as shown in Figures 2g, 2c,2i. The reduction in Chl-*a* has not been observed in Southeast Australia Figure 2h. This could mean that either the productivity in this region has not been affected by the pandemic or the significant drop of SST improved the uptake of $CO_2$ and caused the increase in Chl-*a* and PIC:POC in this region.

The aforementioned reduction of Chl-*a* (50%) during the pandemic is mostly associated with reduction of PIC:POC, which has reached up to 60%. This drop of PIC:POC is observed in most of the regions during the pandemic as evident in Figure xxx. For example, a drop of PIC:POC is observed in Alaska in 2020 for the entire pandemic period. Similarly the reduction of PIC:POC is observed in North Europe in 2020 where PIC:POC doesn't exceed 0.5. There is a prominent reduction in PIC:POC in the northwest Africa with a maximum of 80% in Jun 2020 compared to that of 2019 (Figure 2i). There is also a general reduction of PIC:POC in China & South Korea, which reflect the effect of the $CO_2$ reduction (Figure 2a). In some periods, PIC:POC values are found to be more correlated with SST than those of Chl-*a*. For example, in the southwest Africa, PIC:POC has decreased in the end of April by 60% during the pandemic period of 2020 compared to 2019. This reduction in PIC:POC is correlated with the increase in SST. Likewise, a drop of PIC:POC is observed in the Pacific Ocean during the early pandemic period, April 2020 as shown in Figure 2c. But, later during Jun 2020 PIC:POC values are found to be more than those of 2019 due to a decrease in SST.

The reduction in $CO_2$ emissions has also caused a noticeable drop in SST especially in Alaska, Northeast USA, and Middle East as in Figures 2e,2j,2f. For example, although the general trend of SST in Alaska during the pandemic and pre-pandemic period are similar to each other, the SST during the pandemic year (2020) is found to be slightly lower (by ~0.3 °C) than that of the previous year (2019). In the Southeast of USA, SST shows lower values during the pandemic period compared to the previous year (Figure 2k). Similar drop of SST occurs in China and South Korea in the pandemic period especially in the middle of May and beginning of Jun. The drop of SST in the Middle East is prominent in Jun 2020 compared to that of 2019.

**4. Conclusions**

In conclusion, the COVID-19 pandemic has forced in a 7% reduction of the anthropogenic $CO_2$ emissions in the world. Herein, effect of $CO_2$ reductions on productivity of oceans has been studied in terms of analyzing the changes of Chl-*a*, PIC:POC and SST during the pandemic and pre-pandemic periods. The results show a reduction in Chl-*a* concentrations in the global oceans especially in the coastal regions. For example, there is a prominent decrease in Chl-*a* off Alaska, North Europe, South China and Southeast USA. The drop of mean Chl-*a* by 5% in South China correspond to reduction of 123 $MtCO_2$ in $CO_2$ emissions during the pandemic. The aforementioned reduction of Chl-*a* during the pandemic is mostly associated with reduction of PIC:POC (by 60%). The reduction in $CO_2$ emissions has also caused a drop in SST in Alaska, Northeast USA, Southeast USA and Middle East. The results suggest that a 7% reduction of $CO_2$ emissions within 2 months of lockdown period could reserve the oceans and maintaining similar sustainable practices could highly decrease the climate change affect on oceans.

*14*